\documentstyle[12pt]{article}
\newcommand{\beq}{\begin{equation}}
\newcommand{\eeq}{\end{equation}}
\newcommand{\bea}{\begin{eqnarray}}
\newcommand{\eea}{\end{eqnarray}}
\newcommand{\half}{\frac{1}{2}}
\newcommand{\ihalf}{\frac{i}{2}}

\newcommand{\G}{\Gamma}

\newcommand{\pdp}{\Phi^{\dagger}\Phi}

\newcommand{\dmu}{\partial_{\mu}}

\newcommand{\e}{\epsilon}
\renewcommand{\l}{\lambda}
\newcommand{\oz}{\bar{z}}
\newcommand{\ze}{\oz}

\newcommand{\tres}{\;\;\;}
\begin{document}
\begin{flushright}
La Plata-Th 97/18
\end{flushright}
\vspace{3mm}

\begin{center}

{\large {\bf BPS Solitons and Killing Spinors in}}\\
{\large {\bf Three Dimensional $N=2$ Supergravity}}
\vspace{1cm}

{\bf Jos\'e D. Edelstein}
\vspace{.5cm}

{\it Departamento de F\'{\i}sica, Universidad Nacional de La Plata} \\
{\it C.C. 67, (1900) La Plata, Argentina} \\ 
\vspace{6mm}

{\it Short invited talk presented in} Trends in Theoretical
Physics \\
{\bf CERN--Santiago de Compostela--La Plata Meeting} \\
{\it La Plata, April 28th--May 6th, 1997}
\end{center}

\vspace{5mm}

\begin{abstract}
In the framework of three dimensional extended supergravity 
theories, we demonstrate that there exist non-trivial Killing spinors 
over BPS soliton configurations, even when the space-time is 
asymptotically conical. We also show that there are no physical 
fermionic zero modes on these backgrounds. 
We further generalize these results to the case of semilocal
systems.
\end{abstract}
\vspace{5mm}

The connection between Bogomol'nyi bounds \cite{Bogo} for the mass 
of extended solutions in bosonic field theories and the supercharge 
algebra of the corresponding supersymmetric extensions is by now well 
understood. 
The topological charges of the purely bosonic theory coincide with 
the central extensions of the supersymmetry algebra in the $N$-extended 
model \cite{WO,HS,ENS}. 
The Bogomol'nyi bound then arises from the positivity of the supercharge 
algebra. 
The conditions imposed by the $N$-extended supersymmetric invariance are 
exactly those ensuring the existence of a Bogomol'nyi bound in the purely 
bosonic system. 
Moreover, the bound is saturated by field configurations that preserve a 
certain amount of the supersymmetries of the Lagrangian.
The classical approximation to the mass spectrum 
given by these bounds is expected to be exact at the quantum level as a
consequence of the supersymmetric invariance.

An analogous scenario hold when one includes gravity through {\it local} 
supersymmetry \cite{gibb1,gibb2,BBS,ENS2}. 
As in the global case, some of the supersymmetries are broken when the 
bound is saturated.
The presence of remnant unbroken supersymmetries amounts to the existence 
of spinors that are supercovariantly constants at infinity \cite{hull}.
In three dimensional spacetimes one would expect that these spinors
do not exist for any static massive configuration due to global 
reasons \cite{marc}.
Consider, indeed, the metric of a static spacetime 
\beq
ds^2 = H^2dt^2 - \Omega^2dzd\ze ~,
\label{metric}
\eeq
where we have written the metric on the surface $\Pi$, orthogonal 
everywhere to the time-like killing vector field 
$\frac{\partial}{\partial t}$, in terms of a K\"ahler 
form and complex local coordinates, $\Omega$ being the conformal 
factor. Functions $H$ and $\Omega$ depend only on complex coordinates
$z$ and $\ze$. 
Far from the finite-energy matter sources, it is well-known \cite{DJtH}
that the metric must approach to a cone with deficit angle $\delta$.
This behaviour can be expressed in terms of the following asymptotic 
conditions for functions $H$ and $\Omega$,
\beq
H \to 1 \tres , \tres \Omega \to |z|^{-\delta/\pi} ~.
\label{metricbis}
\eeq
Let us consider for simplicity that $H$ equals $1$ everywhere.
With this metric, the only non-vanishing component of the Einstein 
tensor is $G_{00}$ --the Gauss curvature $K$ of the 
two-dimensional metric $k_{ij}$ that spans $\Pi$-- whose surface
integral can be evaluated using the Gauss-Bonnet theorem:
\beq
\int_{\Pi} dzd\ze \Omega^2 K = \delta - 4\pi g ~,
\label{gabon}
\eeq
where $g$ is the number of handles of the two-dimensional manifold $\Pi$. 
We will be concerned with the simplest case where the topology of $\Pi$
is that of a two-disk, $g = 0$. 
The $00$-component of the Einstein equations can be used to express
the deficit angle in terms of the total mass of the field 
configuration
\beq
\delta =  \frac{1}{M_{pl}} \int_{\Pi} dzd\ze \Omega^2 T^{0~mat}_{~0} =
\frac{M}{M_{pl}} ~.
\label{delta}
\eeq
Now, the parallel transport of a Killing spinor $\eta$ with definite 
`chirality', $\gamma^0\eta_{\pm} = \pm\eta_{\pm}$, 
around a closed curve $\Gamma$ of large radius $R$ --surrounding all
the static matter sources-- is given by the following path-ordered 
integration: 
\beq
\eta_{\pm}(R,2\pi) = {\cal P}\exp\left(-\ihalf\oint_{\Gamma}
\omega_{\mu}^a\gamma_adx^{\mu}\right)\eta_{\pm}(R,0) ~,
\label{paral}              
\eeq
which amounts --for masses below the Planck scale-- to a non-trivial 
holonomy
\beq
\eta_{\pm}(R,2\pi) = \exp\left[\pm i\pi\delta\right]\eta_{\pm}(R,0) 
\label{paral1}
\eeq
that inhibites the existence of well-defined asymptotical Killing spinors.
This state of affairs lead Witten \cite{Witten1} to conjecture
that supersymmetry might ensure the exact vanishing of the 
cosmological constant {\em without} compelling bosons and fermions to 
be degenerate:
Even when supersymmetry applies to the vacuum ensuring the vanishing of 
the cosmological constant, it is broken over massive states as a 
consequence of the geometrical holonomy.

It was recently shown that the obstruction to the existence of 
asymptotical Killing spinors in three-dimensional supergravities can
be eluded for systems admitting of topological solitons 
\cite{BBS,ENS2,ENS3}. In these systems, the geometrical phase 
associated with the conical geometry results to be cancellated by an 
Aharonov--Bohm--like phase that appears in presence of BPS solitons.
An analogous result was found in a new class of $(p,q)$-extended 
Chern-Simons Poincar\'e supergravities studied in Ref.\cite{HIPT}. 

The would-be fermionic Nambu-Goldstone
zero modes generated by the action of broken generators result to be
non-normalizable, thus not entering in the physical Hilbert space
\cite{BBS}. Then, in spite of being possible to end with Killing 
spinors over certain solitonic backgrounds, there is no Bose-Fermi 
degeneracy. It is still possible to have a vanishing cosmological 
constant without implying such a degeneracy in the spectrum.

On general grounds, the argument goes as 
follows \cite{ENS3}. Consider a $2+1$ system of 
gauged matter coupled to $N=2$ supergravity admitting of BPS solitons.
Whenever the system has a non-trivial topologically conserved current 
$J_{\mu}$, $J_{\mu} = \e_{\mu\nu\lambda}\partial^{\nu}{\cal A}^{\l}$, 
the vector potential ${\cal A}_{\mu}$ should belong to a vector 
multiplet, thus effectively contributing to the supercovariant 
derivative $\nabla_{\mu}(\omega)$ as
\beq
\hat{\nabla}_{\mu}(\omega,{\cal A},\ldots) \equiv 
{\nabla}_{\mu}(\omega) + i\frac{\kappa^2}{4}{\cal A}_{\mu} + \ldots 
\label{newcov}
\eeq
where $\ldots$ stands for contributions of the spin-1 component of 
each vector multiplet coupled to the Einstein supermultiplet, which
rapidly vanish at infinity. 
In a sense, the gravitino becomes {\it charged}, though ${\cal A}_\mu$ 
needs not to be a dynamical gauge field \cite{ENS2}.

The contribution that comes from the topological vector potential allows 
the cancelation of the conical holonomy by an Aharonov--Bohm--like phase 
produced when surrounding the solitonic configuration, provided it 
saturates the corresponding Bogomol'nyi bound.
In fact, after its parallel transport around the static matter sources, 
the chiral spinors $\eta_{\pm}$ read
\beq
\eta_{\pm}(R,2\pi) = {\cal P}\exp\left(- \ihalf\oint_{\Gamma}
\omega_{\mu}^a\gamma_adx^{\mu} + i\frac{\kappa^2}{4}
\oint_{\Gamma}{\cal A}_{\mu}dx^{\mu}\right)\eta_{\pm}(R,0) ~.
\label{paral2}
\eeq
Then, as a consequence of the fact that the gravitino has acquired a 
charge, an Aharonov--Bohm--like phase appears, which is proportional to 
the topological charge $T$ of the configuration
\beq                                           
T \equiv \oint_{\Gamma}{\cal A}_{\mu}dx^{\mu} ~.
\label{tete}
\eeq
Thus, for static BPS configurations, the holonomies in (\ref{paral2}) 
cancel each other. Unbroken supercharges can then be defined, in 
principle, for Bogomol'nyi saturated states, in spite of the 
asymptotically conical spacetime geometry. The fermionic 
Nambu--Goldstone zero modes, however, receive an infrared 
divergent contribution that can be traced to come
from the conical singularity.
Then, the would-be unbroken supercharges cannot be represented
into the physical Hilbert space of the theory: There is no
Bose--Fermi degeneracy.

More than an asymptotical supercovariantly constant spinor, the
system has an exact Killing spinor
\beq                                      
\hat{\nabla}_{\mu}(\omega,{\cal A},\ldots)\,\eta_{\pm} = 0 ~,
\label{ndeltacero}
\eeq
which can be thought of as the solution of the Bogomol'nyi equation 
that corresponds to the gravitational field. In fact, the integrability
condition of Eq.(\ref{ndeltacero}),
\beq
[\hat{\nabla}_{\mu}(\omega,{\cal A},\ldots),\hat{\nabla}_{\nu}(\omega,{\cal 
A},\ldots)]\,\eta_{\pm} = 0 ~,
\label{integral}
\eeq
happens to be equivalent to the Einstein equation of the purely
bosonic system over BPS configurations\footnote{It is interesting
to point out that this assertion spoils in the low energy field
theory of heterotic string theory compactified on $T^7$ with a
non-vanishing $3$--form field strength $H_{\mu\nu\l} \neq 0$,
where solutions to the Killing spinor equation do not solve
the equations of motion \cite{BBLC}.}.

This result can be further generalized to the case of semilocal 
solitons, that is, stable extended classical configurations that take 
place in theories with a simply connected vacuum manifold.
The Lagrangian density of such systems have both global and local 
symmetries. The global symmetry is longer than the local one which 
also has to be completely broken. 
This is the case of the semilocal cosmic string first introduced in 
Ref.\cite{VA}, a vortex solution which has been shown to be stable
even though the manifold of minima for the potential energy does not 
contain non-contractible loops. 
The semilocal cosmic string can be coupled to $N=2$ supergravity through 
the dimensional reduction of the standard electroweak model minimally 
coupled to supergravity, by setting the $SU(2)$ gauge coupling constant 
to zero. The dynamics of the bosonic sector is given by
\beq
V^{-1}{\cal L} = \frac{M_{pl}}{2} R - \frac{1}{4} F_{\mu\nu}F^{\mu\nu}
+ \half (D_{\mu}\Phi)^{\dagger}(D^{\mu}\Phi) - \frac{e^2}{8}(\pdp
- v^2)^2 ~,
\label{modelo}
\eeq
where $V$ is the determinant of the dreibein.
The Higgs field $\Phi$ is a complex doublet
and the covariant derivative reflects the fact that we have only
gauged the $U(1)$ factor of the gauge group,
$D_{\mu} = \dmu - ieA_{\mu}$. The Lagrangian (\ref{modelo}) has
a global $SU(2)$ symmetry as well as a local $U(1)$ invariance. 
The vacuum manifold is the
three-sphere $|\Phi| = v$, which has no non-contractible loops. 
However, as the gradient energy density must fall off sufficiently
fast asymptotically, fields at infinity owe to lie on a gauge
orbit, that is, a circle lying on the three-sphere.

The fermionic superpartners receive a charge contribution $\delta q$,
\beq
\delta q = \frac{ev^2}{4M_{pl}} ~,
\eeq
as a consequence of the coupling of our system to $N=2$ supergravity 
with a Fayet-Iliopoulos term. This fact reflects the correction
received by the supercovariant derivative when a vector superfield
is coupled to the Einstein multiplet that was discussed previously.

We inmediately see that the charged spinor $\eta$, acquires
an Aharonov-Bohm phase provided some magnetic flux $\Phi_n$ exists across 
the surface delimited by $\G$. Indeed, eq.(\ref{paral2}) for this
system, takes the form
\beq
\eta_{\pm}(R,2\pi) = \exp\left[ \pm i\pi \left( \delta \pm 
\frac{M_v^2}{4eM_{pl}}\Phi_n \right) \right]\eta_{\pm}(R,0) ~,
\label{pax2}
\eeq
where $M_v^2 = e^2v^2$ is the `photon' mass.
There are vortex configurations that solve the Bogomol'nyi equations
of the system (take, e.g. $\eta_+$)
\beq
{\cal F} = \frac{e}{2}(\Phi^{\dagger}\Phi - v^2) \tres , \tres
D_z\Phi = 0 ~,
\label{bogomas}
\eeq
for which phases exactly cancel \cite{E} (${\cal 
F}$ stands for the magnetic field). Then, asymptotical Killing spinors 
(indeed, exact ones) do exist over these $2+1$-dimensional solitonic
background solutions, even though a na\"{\i}ve analysis of the vacuum 
manifold of this system would lead us to conclude that it has a trivial 
topology. This further 
extends the class of $2+1$-dimensional systems studied in \cite{ENS3}, 
that admits Killing spinors over massive configurations.

Let us attempt to build the entire massive supermultiplet 
associated to the BPS semilocal cosmic string. To this end, we must 
apply the broken supersymmetry generator to our purely bosonic 
configuration. The transformations
with antichiral spinorial parameter $\eta_-$ are nothing but the 
Nambu--Goldstone fermionic zero mode corresponding to
the broken supersymmetry. It is not difficult to see that the
antichiral supersymmetric transformation law of the gravitino,
evaluated on the BPS semilocal vortex solution, 
\beq
\int dzd\ze\Omega^{-1} 
(\delta_{\eta_-}\psi_{\mu})\vert_{BPS}^{\dag}
(\delta_{\eta_-}\psi_{\mu})\vert_{BPS}
\sim \int \frac{dzd\ze}{|z|^2} ~,
\eeq
leads to an infrared divergent contribution that renders the zero mode 
non-normalizable. As a consequence, the zero mode must not be used to 
construct the physical Hilbert space of the theory. 
That is, though there seem to exist unbroken supersymmetries
over certain massive configurations, they cannot be realized on the
physical spectrum. 
Hence, in the semilocal model, the vanishing of the cosmological
constant implied by the supersymmetries of the vacuum, does not compel
bosons and fermions to be degenerate. 

~

This work was partially supported by Consejo Nacional de Investigaciones
Cient\'{\i}ficas y T\'ecnicas (CONICET), Argentina. 
I would like to thank the organizers of the Meeting for their kind
invitation.
I am grateful to Paul Townsend for stimulating discussions.
		  

\end{document}